%% file: main.tex
\documentclass[twocolumn]{aastex63}

\usepackage{apjfonts}
\usepackage{lineno}
\usepackage{xspace}

\newcommand{\glob}{{NGC~3201}\xspace}
\newcommand{\gjoll}{{Gj\"{o}ll}\xspace}
\newcommand{\Gaia}{{\it Gaia}\xspace}

\shorttitle{Linking \gjoll and \glob}
\shortauthors{Hansen et al.}

\begin{document}
\title{A chemo-dynamical link between the \gjoll stream and \glob}

\input{authors.tex}

\correspondingauthor{T.~T.~Hansen}
\email{thansen@tamu.edu}

\graphicspath{{./}{figures/}}

\begin{abstract}
Recent analysis of the second data release of \Gaia has revealed a number of new stellar streams surrounding the Milky Way. It has been suggested that one of these streams, \gjoll, is associated with the globular cluster \glob, but thus far the association has only been based on kinematics of the stream stars. In this work we identify five likely stream members in the \Gaia data that are bright enough for high-resolution spectroscopic follow-up with the Harlan J. Smith telescope at McDonald Observatory. One star is ruled out as a member based on its radial velocity. Abundance analysis of the remaining four kinematic members reveals a good chemical match to \glob for two of the stars, driven by their similar Al and $\alpha$-element abundances. Of the remaining two, one shows no chemical similarity to \glob and is likely an unassociated Milky Way halo field star, while the other exhibits a similar Al abundance but has somewhat lower $\alpha$-element abundances. The chemical tagging of stars in the \gjoll stream to \glob confirms this association and provides direct proof of the hierarchical formation of the Milky Way.
\end{abstract}


\section{Introduction \label{sec:intro}}
Stellar streams, the remnants of accreted globular clusters (GCs) and dwarf galaxies, are excellent probes of galaxy formation and cosmology. They are natural consequences of hierarchical galaxy formation, forming as their progenitors fall into and are unraveled by tidal forces of their host galaxy. Observations of Milky Way streams can be used to constrain the formation of the stellar halo \citep{johnston1998,bullock2005}, the total mass and shape of the Galactic potential \citep{koposov2010,law2010, pricewhelan2014,bonaca2014,erkal2016,bovy2016,malhan2019}, and perturbations in the gravitational field due to known satellites \citep{gomez2015,erkal2018,erkal2019} and low-mass dark matter substructures \citep{ibata2002,johnston2002,carlberg2009,carlberg2012,erkal2015,banik2019}. Wide-area digital photometric surveys have dramatically increased the number of known streams around the Milky Way (e.g.~\citealt{shipp2018}; see \citealt{riley2020} for a recent compilation) and other nearby galaxies \citep{martinez2010}, possibly leading to a new era of statistical modeling of the stream population \citep{bonaca2018}.

The second public release of \Gaia data \citep[DR2;][]{dr2brown2018} has also contributed to this renaissance of stream data.
Several studies have used \Gaia DR2 proper motions to remove foreground contamination around known streams and reveal new features \citep{malhan2018a,malhan2019a,bonaca2019a,bonaca2019b,bonaca2020} or characterize the stream's dynamics \citep{shipp2019, pricewhelan2019,ibata2020}. In addition, over 15 new streams have been discovered utilizing this dataset, mostly by using the \texttt{STREAMFINDER} algorithm \citep{malhan2018b,ibata2018,ibata2019b,malhan2019b}. Many of these streams have been kinematically confirmed with independent radial velocity measurements of the stream stars \citep{ibata2019b,roederer2019,malhan2019b}, proving that \Gaia proper motions can be used to uncover low surface brightness streams that are missed in photometric studies.

\citet{ibata2019b} linked one of these new streams, Fimbulthul, to the massive GC $\omega$ Cen. They showed that the two shared similar orbital energies and angular momenta. \citet{ibata2019a} expanded on this finding, using N-body simulations to show that $\omega$ Cen's expected tidal tail overlaps with Fimbulthul and high-resolution spectroscopy of five stars in the stream to show that the radial velocities and metallicities are consistent with the stars having originated in the cluster.

GCs are known to display very uniform metallicities and prominent element abundance correlations and anti-correlations. The most well-known of these is the Na-O anti-correlations detected in most of the GCs analyzed today \citep[e.g.][]{carretta2009a}. This abundance signature is believed to be the result of material being processed through the CNO and Ne-Na cycles in the first generation of stars in the cluster. However, the Mg-Al cycle was likely also active in the polluters of GC stars resulting in specific trends in the Mg, Al, and possibly Si abundances of the second generation stars in GCs \citep{yong2015}. These distinct chemical signatures makes GC stars stripped from their original cluster easily identifiable. In fact, \citet{simpson2020} used elemental abundances from the GALAH survey \citep{desilva2015,buder2018} to chemically tag two Fimbulthul stars to $\omega$ Cen, strengthening the link between the two structures. But the \citet{simpson2020} study also rejected two kinematically-selected stars based on their detailed chemical abundances, highlighting the need for a combination of kinematic and detailed chemical analysis to tie streams to their parent system.

In this work, we aim to use these techniques to chemo-dynamically link another cluster-stream pairing: \glob and the \gjoll stream. \gjoll was discovered by \citet{ibata2019b}, who applied the \texttt{STREAMFINDER} algorithm to \Gaia DR2 data in the inner Galaxy. Both the stream \citep{ibata2019b} and the cluster \citep{dr2helmi2018} are on strongly retrograde orbits ($L_z \sim 2700$~kpc~km~s$^{-1}$) with similar pericenters ($\sim$8~kpc) and apocenters ($\sim$30~kpc). Noting this alignment in phase space, \citet{bianchini2019} used \Gaia DR2 data to identify tidal tails coming off of the cluster. Furthermore, orbit integrations of \glob pass through both endpoints of the \gjoll stream \citep{riley2020}. Taken together, this evidence suggests that \glob and \gjoll are dynamically linked; establishing a chemical link between the two would solidify this association.

This work is structured as follows. In Section \ref{sec:orbit} we present the orbit integration evidence that dynamically links \glob to \gjoll and use this orbit to select target stars for high-resolution spectroscopic follow-up. Sections \ref{sec:obs} and \ref{sec:param} describe the high-resolution observation and chemical analysis of the kinematically-selected candidates. Using those results, we discuss the chemical membership of the stars in Section \ref{sec:results} and provide a summary in Section \ref{sec:summary}.

\begin{figure*}[hbt!]
\centering
\includegraphics[width=\textwidth]{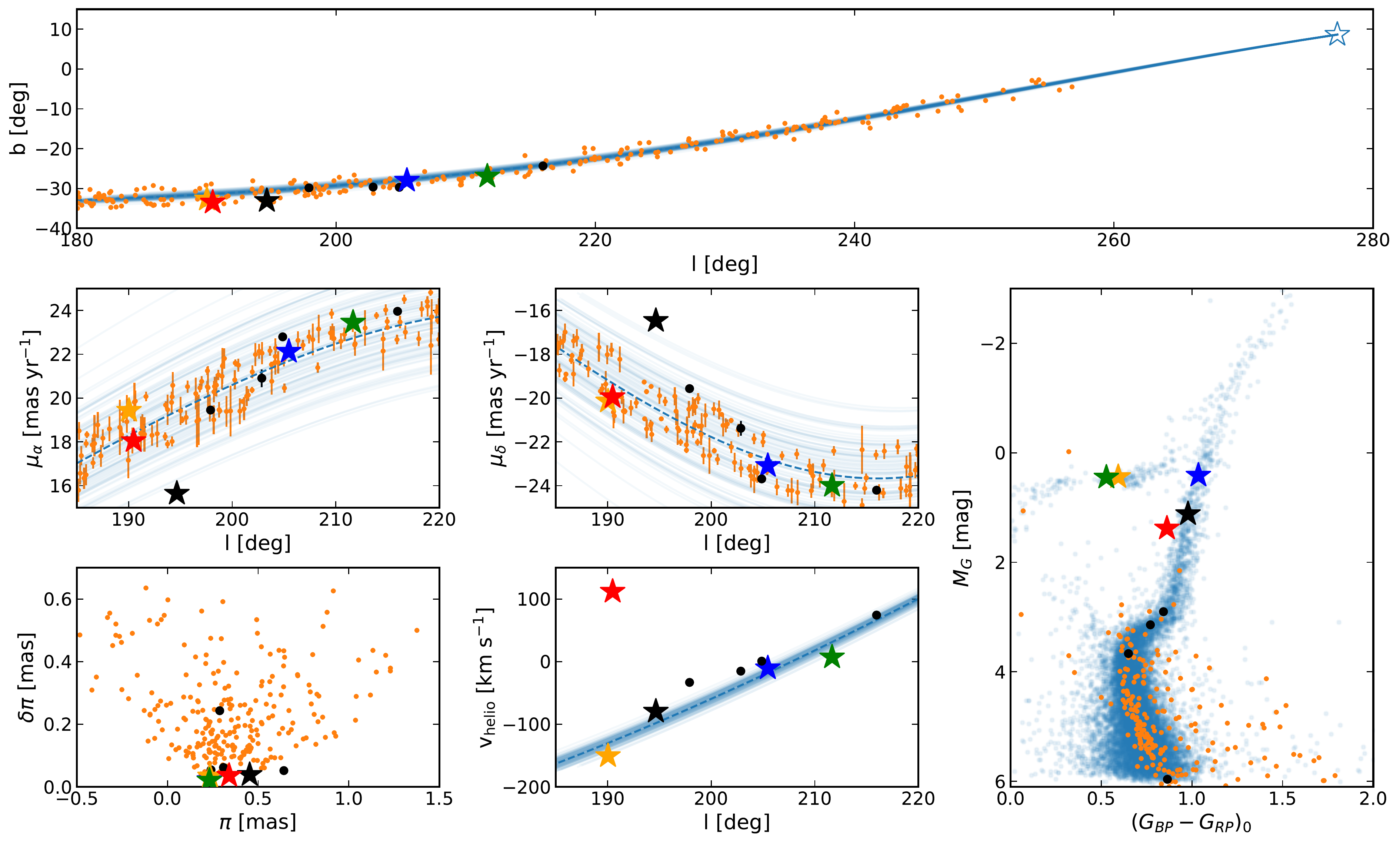}
\caption{Orbit of \glob through \gjoll and kinematically-selected target stars. Each of the 100 faint blue lines \textbf{originating from \glob's present-day position (large hollow star)} corresponds to a Monte Carlo orbit in the \citet{pricewhelan2019} potential. The dashed blue line corresponds to the fiducial orbit (assuming no measurement errors). Orange points (with errors) correspond to stars that have 5-D kinematic information that is consistent with \glob's orbit (see text for details). The bottom-right panel is a color-magnitude diagram with blue points for \glob from \citet{dr2helmi2018}. Each of the large, colored stars were followed up spectroscopically with McDonald. Small black circles correspond to \gjoll members identified by \citet{ibata2019b} that were too faint for follow up. \label{fig:orbit}}
\end{figure*}

\section{Orbit Integration and Target Selection} \label{sec:orbit}
In an effort to associate Milky Way satellite galaxies and globular clusters with known stellar streams, \citet{riley2020} integrated orbits for each satellite to see if they passed through both endpoints for any stream.
They accounted for observational errors using Monte Carlo simulations and repeated the procedure for three different Milky Way potentials \citep{law2010,bovy2015,pricewhelan2017}.
Regardless of which potential was assumed, \glob had a high fraction of Monte Carlo orbits pass through both endpoints of the \gjoll stream (30-60\%, while most satellite-stream pairings had zero).
\glob and \gjoll are also located relatively close to each other in physical space; integrating the cluster's orbit backwards $\sim$20 Myr overlaps with the stream.

Given this likely association, we targeted stars that were probable members of \gjoll for spectroscopic follow-up.
We selected stars from \Gaia DR2 based on the following criteria relative to \glob's orbit in the potential from \citet{pricewhelan2017}:
\begin{itemize}
    \item[--] Position on the sky within 3 degrees of the orbit trajectory (excluding a region of 2 degrees around the cluster's present location)
    \item[--] Measured parallax $|\pi - 1/D| < 3\epsilon_\pi$, where $D$ is the heliocentric distance of the orbit and $\epsilon_\pi$ is the quoted parallax uncertainty
    \item[--] Measured proper motions $(\mu_{\alpha\cos\delta}, \mu_\delta)$ each within 1.5 mas yr$^{-1}$ of the orbit's proper motion
\end{itemize}

We also removed stars whose astrometric fits are potentially unreliable.
As detailed in \citet{lindegren2018}, we remove sources that have \texttt{astrometric\_excess\_noise} greater than 1 or re-normalized unit weight error $u > 1.2\times \max(1, \exp(-0.2(G-19.5))$.
We also apply the following recommended cut to remove stars with significant color excess $E$ (\texttt{phot\_bp\_rp\_excess\_factor}): 1.0 + 0.015 \texttt{bp\_rp}$^2$ < $E$ < 1.3 + 0.06 \texttt{bp\_rp}$^2$.
Finally, we include any stars that were identified as \gjoll members by \citet{ibata2019b} in their Tables 1 and 3.

While we do not use photometry to select target stars, we did inspect the color-magnitude diagram (CMD) of our kinematically-selected stars and compared to the CMD for stars selected to be \glob members by \citet{dr2helmi2018}.
The CMD was extinction-corrected using \citet{schlafly2011} corrections to the \citet{schlegel1998} extinction maps, assuming the extinction ratios $A_G/A_V$ = 0.85926, $A_{G_{\text{BP}}}/A_V$ = 1.06794, and $A_{G_{\text{RP}}}/A_V$ = 0.65199, as listed on the web interface to the PARSEC isochrones \citep{bressan2012}.
We also convert to absolute magnitudes to account for the varying heliocentric distance along the orbit, assuming that the stars have a distance that matches perfectly with the orbit distance at that Galactic longitude rather than inverting the measured parallax (see discussion in \citealt{bailerjones2018}).

The results of these selections are illustrated in Figure \ref{fig:orbit}.
Kinematically-selected target stars trace the orbit of \glob over 50 degrees of the sky, overlapping with identified \gjoll members over $195 < l < 215$.
We note that there are no candidate members with $|b| < 10$, likely due to obscuration from the Milky Way. We compile the data of targets in Table \ref{tab:targets}. We conducted a thorough search to identify additional targets for which high-resolution spectra could be obtained, but even significantly loosening the above selection criteria did not produce suitable targets.

While it is possible that some of our candidates are chance kinematic alignments and not former members of \glob, photometric data largely support the association. The vast majority of stars that are kinematically selected to match \gjoll (using \glob's orbit) -- with no photometric cuts -- also preferentially match \glob's CMD (see bottom-right panel of Figure \ref{fig:orbit}). Furthermore, all six stars that were spectroscopically identified as \gjoll members in \citet{ibata2019b} fall along the same CMD (four black points, black star, and blue star in Figure \ref{fig:orbit}). Repeating this procedure with another potential stream-cluster pairing identified by \citet{riley2020}, NGC 4590 and PS1-E, resulted in far fewer kinematic candidates that did not cluster along NGC 4590's CMD. These results suggest that many, if not most, of the candidate stars identified here are part of this disrupting system.

\begin{deluxetable*}{lcrcccccrc}[hbt!]
\tablecaption{Known and suspected \gjoll stream stars \label{tab:targets}} 
\tablewidth{0pt}
\tablehead{\Gaia Source ID & RA & Dec & $G_0$ & $(G_{\text{BP}} - G_{\text{RP}})_0$ & $\pi$ & $\mu_{\alpha\cos\delta}$ & $\mu_\delta$ & $v_{\text{hel}}$ & I+19 \\
& (deg) & (deg) & (mag) & (mag) & (mas) & (mas yr$^{-1}$) & (mas yr$^{-1}$) & (km s$^{-1}$) & }
\startdata
\input{tables/candidates.tex}
\enddata
\tablecomments{The table is separated into stars that have spectroscopic followup (above line) from either this work or \citet{ibata2019b} and those that do not (below line). The final column lists whether the star is part of the more conservative sample from \citet{ibata2019b}. An asterisk in the v$_{\text{hel}}$ column indicates the radial velocity is from \citet{ibata2019b}. The complete version of Table \ref{tab:targets} is available online only.}
\end{deluxetable*}

\begin{deluxetable}{lcccr}[hbt!]
\tablecaption{Observing log\label{tab:obslog}} 
\tablewidth{0pt}
\tablehead{Object Name & Date & $t_{exp}$ & $\rm{SNR}$ &$V_{hel}$ \\
&(HJD) &  (sec)  & $@$5000\AA\   &km~s$^{-1}$ }
\startdata
$Gaia$~32541125 & 2458801.78418 & 7200 &    &$-$79.1$\pm$0.9\\ 
                & 2458831.68457 & 7200 & 30 &$-$80.8$\pm$0.4\\ 
$Gaia$~31873905 & 2458801.87109 & 7200 &    &$-$9.9$\pm$0.9 \\ 
                & 2458831.77246 & 7200 & 29 & $-$10.2$\pm$0.6\\ 
$Gaia$~29901421$^*$ & 2458802.84375 & 7200 &    & $+$7.4$\pm$0.6 \\ 
                & 2458866.66113 & 2700 & 20 & $+$20.1$\pm$1.1\\
$Gaia$~32591587$^*$ & 2458832.75000 & 2400 &    & $-$154.3$\pm$1.2\\
                & 2458888.57422 & 7200 & 20 & $-$144.4$\pm$4.9\\
$Gaia$~32589760 & 2458833.69043 & 2400 & 8  & $+$110.8$\pm$3.0\\ 
\enddata
\tablecomments{$^*$ Date and $V_{hel}$ given for first exposure; see Table \ref{tab:RRL} for details. SNR is measured in the combined spectrum.}
\end{deluxetable}

\input{tables/RRL.tab}

\section{Observations \label{sec:obs}}
The kinematically-selected targets brighter than $G=15$ mag were observed with the Harlan J. Smith 107-in (2.7 m) telescope and the TS23 echelle spectrograph \citep{tull1995} at McDonald Observatory. The spectra were obtained using the 1.8" slit and 1x1 binning yielding a resolving power of $R = \lambda / \Delta \lambda \sim 35,000$, with full wavelength coverage over 3600 $< \lambda <$ 5800 \AA\ and partial wavelength coverage up to 10,000 \AA. The five stars; $Gaia$~3254112556278356608 ($Gaia$~32541125), $Gaia$~3187390548572555136 ($Gaia$~31873905), $Gaia$~2990142148280216960 ($Gaia$~29901421), $Gaia$~3259158764894232192 ($Gaia$~32591587) and $Gaia$~3258976074166599680 ($Gaia$~32589760) were observed over four separate runs from November 2019 to February 2020. Table \ref{tab:obslog} lists the observations of all stars. Two of the stars; $Gaia$~29901421 and $Gaia$~32591587 are RR-Lyrae (RRL) stars for these phase and radial velocities for the individual observations are listed in Table \ref{tab:RRL}. The data were reduced using standard IRAF packages, including correction for bias, flatfield, and scattered light. Signal to noise ratios (SNR) measured at 5000\AA~for the final spectra are listed in Table \ref{tab:obslog}. Heliocentric radial velocities for the stars were determined via cross-correlation with spectra of two radial velocity standard stars, observed on the same nights as the target stars, HD~38230 ($V_{helio}=-29.07$ \citep{soubiran2018}) and HD~122563 ($V_{helio}=-26.17$ \citep{soubiran2018}). Between three and twenty orders in each spectrum were used for the correlation. Table \ref{tab:obslog} lists the mean heliocentric radial velocity and standard deviation for the stars. 
The measured radial velocities rejected $Gaia$~32589760 (red star in Figure \ref{fig:orbit}) as a member while confirming the kinematic membership for the other four stars.

\begin{deluxetable}{lccccc}
\caption{Stellar Parameters\label{tab:params}}
\tablehead{\colhead{ID} & \colhead{$T_{\rm eff}$}& \colhead{$\log g$} & \colhead{$\xi$}  & \colhead{$\mathrm{[Fe/H]}$} \\
&(K)&(cgs)&(km~s$^{-1}$)&(dex)}
\startdata
$Gaia$~32541125 & 5250$\pm$ 100& 3.55$\pm$0.3& 0.50$\pm$0.3& -1.03$\pm$0.17\\ 
$Gaia$~31873905 & 5060$\pm$ 100& 2.48$\pm$0.3& 1.50$\pm$0.3& -1.34$\pm$0.11\\ 
$Gaia$~29901421 & 6300$\pm$ 150& 1.50$\pm$0.3& 2.65$\pm$0.3& -1.68$\pm$0.17\\ 
$Gaia$~32591587 & 6000$\pm$ 150& 2.15$\pm$0.3& 2.00$\pm$0.3& -1.59$\pm$0.21\\ 
\enddata
\end{deluxetable}

\section{Stellar Parameter Determination and Abundance Analysis \label{sec:param}}
Stellar parameter and abundances were derived using the 2017 version of MOOG \citep{sneden1973} and making the assumption of local thermodynamic equilibrium (LTE) and including Rayleigh scattering treatment as described by \citet{sobeck2011}\footnote{https://github.com/alexji/moog17scat}. The stellar parameters for the three stars were determined spectroscopically from equivalent width (EW) measurements of \ion{Fe}{1} and \ion{Fe}{2} lines. EWs were measured by fitting Gaussian profiles to the absorption lines in the continuum-normalized spectra. Uncertainties on the EWs were computed using $\sigma_{EW}=1.5\sqrt{FWHM*\delta x}/\rm{SNR}$ from \citet{cayrel1988}, where SNR is the signal to noise per pixel and $\delta x$ is the pixel size. The effective temperatures were determined from excitation equilibrium of \ion{Fe}{1} lines and surface gravities ($\log g$) were determined from ionization equilibrium between the \ion{Fe}{1} and \ion{Fe}{2} lines. Finally, microturbulent velocities ($\xi$) were determined by removing any trend in line abundances with reduced EW for the \ion{Fe}{1} lines. Final stellar parameters, along with estimated uncertainties, are presented in Table \ref{tab:params}, and lines used for the parameter determination of each star are listed in Table \ref{tab:Fe}. Uncertainties on $T_{\rm eff}$ are estimated by visually inspecting the trend of abundances with excitation potential at varying temperatures. The 100 and 150K uncertainties correspond to trends resulting in 0.2~dex differences for the high and low excitation potential lines. While estimates of the uncertainties for $\log g$ and $\xi$ were determined by examining the combined effect of the standard deviation of the Fe~I abundances and the uncertainty in $T_{\rm eff}$ on these. For the two warm RRL stars significantly fewer \ion{Fe}{1} lines were measurable in the spectra resulting in a higher uncertainty on the temperatures of these.

For $Gaia$~32541125 and $Gaia$~31873905 we also determine photometric temperatures using $V-K$ colors and the temperature scale of \citet{alonso1999}. Photometric transformations from \citet{evans2018} was used to convert the \textit{Gaia} $G_0$, $(G_{BP}-G_{RP})_0$ magnitudes to $V_0$ and $K_0$ magnitudes. This results in $T_{\rm eff,photo}$ $=5218\pm124$K and $5057\pm118$K for $Gaia$~32541125 and $Gaia$~31873905 respectively. These are in good agreement with the temperatures derived from \ion{Fe}{1} listed in Table \ref{tab:params}

Abundances are derived via equivalent width and spectral synthesis analysis using 1D LTE ATLAS9 model atmospheres \citep{castelli2003} and the solar photosphere abundances from \cite{asplund2009}. Table \ref{tab:lineabun} lists the abundances derived from individual lines in each star. Line lists were generated using the linemake package\footnote{https://github.com/vmplacco/linemake}, including molecular lines for CH, C$_2$, and CN, and hyperfine structure information. Isotopic fractions for Ba are from \citet{gallagher2010} anf from \citet{lawler2001} for Eu. Uncertainties on the derived abundances arising from stellar parameter uncertainties were determined including covariance terms following \citet{mcwilliam2013} and \citet{johnson2002}.
The covariances were computed using the following equation
\begin{equation}
\sigma_{XY}=\frac{1}{N}\sum_i^N(X_i-\bar{X})(Y_i-\bar{Y})    
\end{equation}{}
$\sigma_{T\log g}$, $\sigma_{T\xi}$, and $\sigma_{T\mathrm{[M/H]}}$ were determined by generating 20 model atmospheres with effective temperatures drawn from a Gaussian distribution with a mean equal to the $T_{\rm eff}$ of the star and standard deviation equal to the uncertainty on $T_{\rm eff}$. $\log g$ and $\xi$ were then varied in turn until ionization equilibrium between the \ion{Fe}{1} and \ion{Fe}{2} lines was obtained for $\sigma_{T\log g}$ and no trend was visible in line abundances with reduced EW of \ion{Fe}{1} lines for $\sigma_{T\xi}$. For $\sigma_{T\mathrm{[M/H]}}$ the direct chance in $\mathrm{[Fe/H]}$ was used. Similarly to determine $\sigma_{\log g\xi}$, 20 model atmospheres with microturbulences drawn from a Gaussian distribution with a mean equal to the $\xi$ of the star and standard deviation equal to the uncertainty on $\xi$ were computed. The gravity was then again varied to obtain ionization equilibrium between the \ion{Fe}{1} and \ion{Fe}{2} lines. The final covariances resulting from this process are $\sigma_{T\log g}$=16, $\sigma_{T\xi}$=6, $\sigma_{T\mathrm{[M/H]}}$=7, and $\sigma_{\log g\xi}$=-0.1 for $Gaia$~31873905 and $\sigma_{T\log g}$=43, $\sigma_{T\xi}$=0.2, $\sigma_{T\mathrm{[M/H]}}$=14, and $\sigma_{\log g\xi}$=-0.04 for $Gaia$~29901421. Tables \ref{tab:error} and \ref{tab:errorRRL} list uncertainties arising from stellar parameter uncertainties for $Gaia$~31873905 and $Gaia$~29901421, respectively. These were determined by deriving abundances for each star using different atmospheric models, each with one parameter varied by its uncertainty and added in quadrature including covariance terms to provide the systematic uncertainty on $\mathrm{[X/H]}$, $\sigma_{\rm sys}$.

\begin{deluxetable}{llrrrrrr}
\tabletypesize{\scriptsize}
\tablecaption{EW and atomic data for \ion{Fe}{1} and \ion{Fe}{2} lines used for parameter determination. \label{tab:Fe}}
\tablewidth{0pt}
\tablehead{\colhead{Stellar ID} & \colhead{Species} & \colhead{$\lambda$} &
\colhead{$\chi$} & \colhead{$\log$ $gf$} & \colhead{EW} &
\colhead{$\sigma_{\rm EW}$} & \colhead{$\log \epsilon$}\\ 
\colhead{} & \colhead{} & \colhead{(\AA)} & \colhead{(eV)} & \colhead{} &
\colhead{(m\AA)} & \colhead{(m\AA)} & \colhead{}}
\startdata
\input{tables/Fe_ew_stub.tab}
\enddata
\tablecomments{
The complete version of Table~\ref{tab:Fe} is available online only.
A short version is shown here to illustrate its form and content.
}							       
\end{deluxetable}						       

\begin{deluxetable*}{lrrrrrrrrrrr}
\tablecaption{Individual line abundances.\label{tab:lineabun}}
\tablewidth{0pt}
\tabletypesize{\scriptsize}
\tablehead{
\colhead{}& & & &\multicolumn{2}{c}{$Gaia$~32541125} &\multicolumn{2}{c}{$Gaia$~31873905} &\multicolumn{2}{c}{$Gaia$~29901421} &\multicolumn{2}{c}{$Gaia$~32591587}\\
\colhead{Species} & \colhead{$\lambda$} & \colhead{$\chi$} & \colhead{$\log gf$} & \colhead{EW} & \colhead{$\log\epsilon$} & \colhead{EW} &\colhead{$\log\epsilon$} & \colhead{EW} & \colhead{$\log\epsilon$} & \colhead{EW} & \colhead{$\log\epsilon$}\\ 
\colhead{} &\colhead{(\AA)} & \colhead{(eV)} & \colhead{} & \colhead{(m\AA)} & \colhead{(dex)} & \colhead{(m\AA)} & \colhead{(dex)} & \colhead{(m\AA)} & \colhead{(dex)} &\colhead{(m\AA)} & \colhead{(dex)}}
\startdata
\input{tables/lineabun_stub.tab}

\enddata
\tablecomments{
The complete version of Table~\ref{tab:lineabun} is available online only.
A short version is shown here to illustrate its form and content.
}
\end{deluxetable*}

\begin{deluxetable}{lccccc}
\tablecaption{Abundance errors arising from stellar parameter
uncertainties for $Gaia$~31873905.\label{tab:error}} 
\tablehead{\colhead{Element}   & \colhead{$\Delta$T$_{\rm eff}$}   &
\colhead{$\Delta\log g$}   &   \colhead{$\Delta\xi$}   &
\colhead{$\Delta\mathrm{[M/H]}$}   &   \colhead{$\sigma_{\rm sys}$}\\
\colhead{} &  \colhead{(dex)}   & \colhead{(dex)}   &   \colhead{(dex)}   &  \colhead{(dex)}   &   \colhead{(dex)} }
\startdata
\input{tables/error.tab}

\enddata
\tablecomments{$^*$ Derived for $Gaia$~32541125}
\end{deluxetable} 

\begin{deluxetable}{lccccc}
\tablecaption{Abundance errors arising from stellar parameter
uncertainties for $Gaia$~29901421.\label{tab:errorRRL}} 
\tablehead{\colhead{Element}   & \colhead{$\Delta$T$_{\rm eff}$}   &
\colhead{$\Delta\log g$}   &   \colhead{$\Delta\xi$}   &
\colhead{$\Delta\mathrm{[M/H]}$}   &   \colhead{$\sigma_{\rm sys}$}\\
\colhead{} &  \colhead{(dex)}   & \colhead{(dex)}   &   \colhead{(dex)}   &  \colhead{(dex)}   &   \colhead{(dex)} }
\startdata
\input{tables/error_RRL.tab}

\enddata
\end{deluxetable} 

\section{Abundance Results \label{sec:results}}

Abundances or upper limits have been derived for 24 elements from C to Eu in $Gaia$~32541125, $Gaia$~31873905, $Gaia$~29901421, and $Gaia$~32591587. All abundances and upper limits are presented in Table \ref{tab:abun1} and \ref{tab:abun2}, listing the $\log_\epsilon$ (X) abundances, the number of lines used to derived the abundance, standard deviation ($\sigma_{\rm stat}$) along with $\mathrm{[X/H]}$ and $\mathrm{[X/Fe]}$ and associated uncertainties on these. For elements where only one or two line was used to derive the abundance we use an estimated $\sigma_{\rm stat}=0.2$ based on the standard deviation for elements with more lines available.

The four stars display somewhat different abundances. Most of them exhibit a mild enhancement in their $\alpha$-element abundances, with the exception of $Gaia$~32591587 which is $\alpha$-poor. Also, $Gaia$~32541125 displays an overall enhancement in neutron-capture elements, allowing us to derive abundances for a number of neutron-capture elements in this star. We find a barium to europium ratio of $\mathrm{[Ba/Eu]}= 0.14$ in this star suggesting a mixed $s$- and $r$-process origin of the neutron-capture elements in this star \citep[e.g.][]{frebel2018}. 

\begin{deluxetable*}{lrrrrrrrrrrrrrr}
\tabletypesize{\scriptsize}
\tablecaption{Derived abundances\label{tab:abun1}}
\tablewidth{0pt}
\tablehead{
\colhead{}&\multicolumn{7}{c}{$Gaia$~31873905} & \multicolumn{7}{c}{$Gaia$~29901421}\\
\colhead{X} & \colhead{$\log\epsilon$(X)} & \colhead{N} &
\colhead{$\sigma_{\rm stat}$}& \colhead{$\mathrm{[X/H]}$} &
\colhead{$\sigma_\mathrm{[X/H]}$} & \colhead{$\mathrm{[X/Fe]}$} & \colhead{$\sigma_\mathrm{[X/Fe]}$}
& \colhead{$\log\epsilon$(X)} & \colhead{N} & \colhead{$\sigma_{\rm stat}$} & \colhead{$\mathrm{[X/H]}$}  & \colhead{$\sigma_\mathrm{[X/H]}$} & \colhead{$\mathrm{[X/Fe]}$} & \colhead{$\sigma_\mathrm{[X/Fe]}$}\\
\colhead{} &\colhead{(dex)} &\colhead{} &\colhead{(dex)} &\colhead{(dex)}
&\colhead{(dex)} &\colhead{(dex)} &\colhead{(dex)} &\colhead{(dex)}
&\colhead{} &\colhead{(dex)} &\colhead{(dex)} &\colhead{(dex)} 
&\colhead{(dex)} &\colhead{(dex)}}
\startdata
\input{tables/abun1.tab}

\enddata
\end{deluxetable*}

\begin{deluxetable*}{lrrrrrrrrrrrrrr}
\tabletypesize{\scriptsize}
\tablecaption{Derived abundances\label{tab:abun2}}
\tablewidth{0pt}
\tablehead{
\colhead{}&\multicolumn{7}{c}{$Gaia$~32591587} & \multicolumn{7}{c}{$Gaia$~32541125}\\
\colhead{X} & \colhead{$\log\epsilon$(X)} & \colhead{N} & \colhead{$\sigma_{\rm stat}$}
&\colhead{$\mathrm{[X/H]}$}  & \colhead{$\sigma_\mathrm{[X/H]}$} & \colhead{$\mathrm{[X/Fe]}$}  & \colhead{$\sigma_\mathrm{[X/Fe]}$}
& \colhead{$\log\epsilon$(X)} & \colhead{N} & \colhead{$\sigma_{\rm stat}$} &\colhead{$\mathrm{[X/H]}$}  & \colhead{$\sigma_\mathrm{[X/H]}$} & \colhead{$\mathrm{[X/Fe]}$}  & \colhead{$\sigma_\mathrm{[X/Fe]}$}\\
\colhead{} & \colhead{(dex)} &\colhead{} & \colhead{(dex)} & \colhead{(dex)}
&\colhead{(dex)} & \colhead{(dex)} &\colhead{(dex)} & \colhead{(dex)} &\colhead{} & \colhead{(dex)} & \colhead{(dex)}
&\colhead{(dex)} & \colhead{(dex)} &\colhead{(dex)}}
\startdata
\input{tables/abun2.tab}

\enddata
\end{deluxetable*}

\begin{figure*}[hbt!]
\centering
\includegraphics[width=0.7\textwidth]{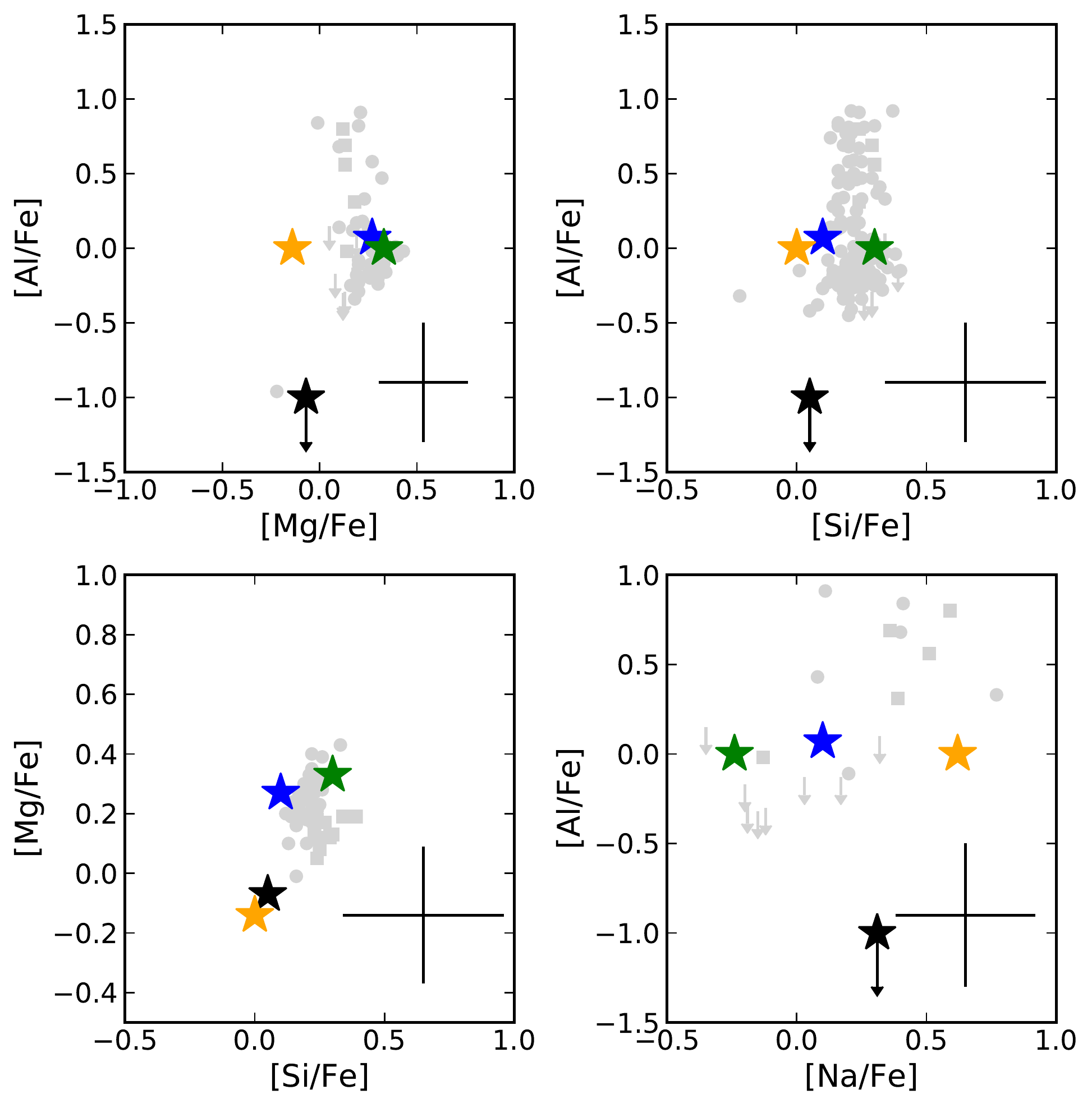}
\caption{Combination of $\mathrm{[Na/Fe]}$, $\mathrm{[Mg/Fe]}$, $\mathrm{[Al/Fe]}$, and $\mathrm{[Si/Fe]}$ abundances for $Gaia$~32541125 (black star), $Gaia$~31873905 (blue star), $Gaia$~29901421 (green star), and $Gaia$~32591587 (yellow star). Literature abundances for stars in \glob are plotted in grey. Dots are abundances taken from \citet{meszaros2020} while squares and triangles are abundances and upper limits, respectively, from \citet{carretta2009b}. Abundances from \citet{carretta2009b} are used for stars overlapping between the two samples. A representative error bar for the \gjoll abundances is shown in the lower right corner.
\label{fig:abun}}
\end{figure*}

\subsection{Chemical Membership}
As describe in the Introduction, the abundances for GC stars generally show distinct patterns, distinguishing them from other stellar populations. \glob is no exception: it has, like many other clusters, been found to have a very uniform metallicity. \citet{carretta2009a} find $\mathrm{[Fe/H]} = -1.50$ with a very small intrinsic scatter of $\sigma=0.05$ for their sample of 149 stars while \citet{meszaros2020} find a somewhat higher value of $\mathrm{[Fe/H]}=-1.24$ with a scatter of $\sigma =0.10$ from their sample of 179 stars. Part of this discrepancy is due to the difference in solar Fe abundances used, \citet{carretta2009a} used A(Fe)$_{\odot}$=7.54 \citep{gratton2003} while \citet{meszaros2020} used A(Fe)$_{\odot}$=7.45 \citep{grevesse2007}. However, an offset persists, even after correction for the different A(Fe)$_{\odot}$ values, which is likely due to a systematic difference in the effective temperatures used for the two studies. Both studies use photometric temperatures but with different scales \citep{meszaros2020}. Other recent work analysing smaller samples of stars in \glob find mean metallicities similar to \citet{carretta2009a}, e.g. $\mathrm{[Fe/H]}=-1.53$ \citep{munoz2013}, $\mathrm{[Fe/H]}=-1.48$ \citep{simmerer2013} ,  $\mathrm{[Fe/H]}=-1.42$ \citep{mucciarelli2015}, and $\mathrm{[Fe/H]}=-1.47$ \citep{magurno2018}. In this work we find a spread in the metallicities derived for the four \gjoll stars from $\mathrm{[Fe/H]}= -1.68$ to $-1.03$. Our parameters are purely spectroscopically derived, however, for $Gaia$~32541125 and $Gaia$~31873905 we also derived photometric effective temperatures following a similar approach as \cite{carretta2009b}. Our photometric and spectroscopic temperatures are in good agreement, suggesting our temperature scale is more similar to \citet{carretta2009b} than \citep{meszaros2020}. Furthermore, \cite{carretta2009a} corrected their EWs measured from the intermediate resolution GIRAFFE spectra to a system defined by the high-resolution UVES spectra to account for unrecognized blends in the GIRAFFE EWs resulting in an optical EW analysis similar to the one presented in this paper. In any case, the large spread in our metallicities suggests that no all stars where stripped from \glob, and highlights the need for a multi-element abundance analysis of the stars, as presented here, to access the actual connection between a stream's stars and its parent object.

\begin{figure}[hbt!]
\centering
\includegraphics[width=\linewidth]{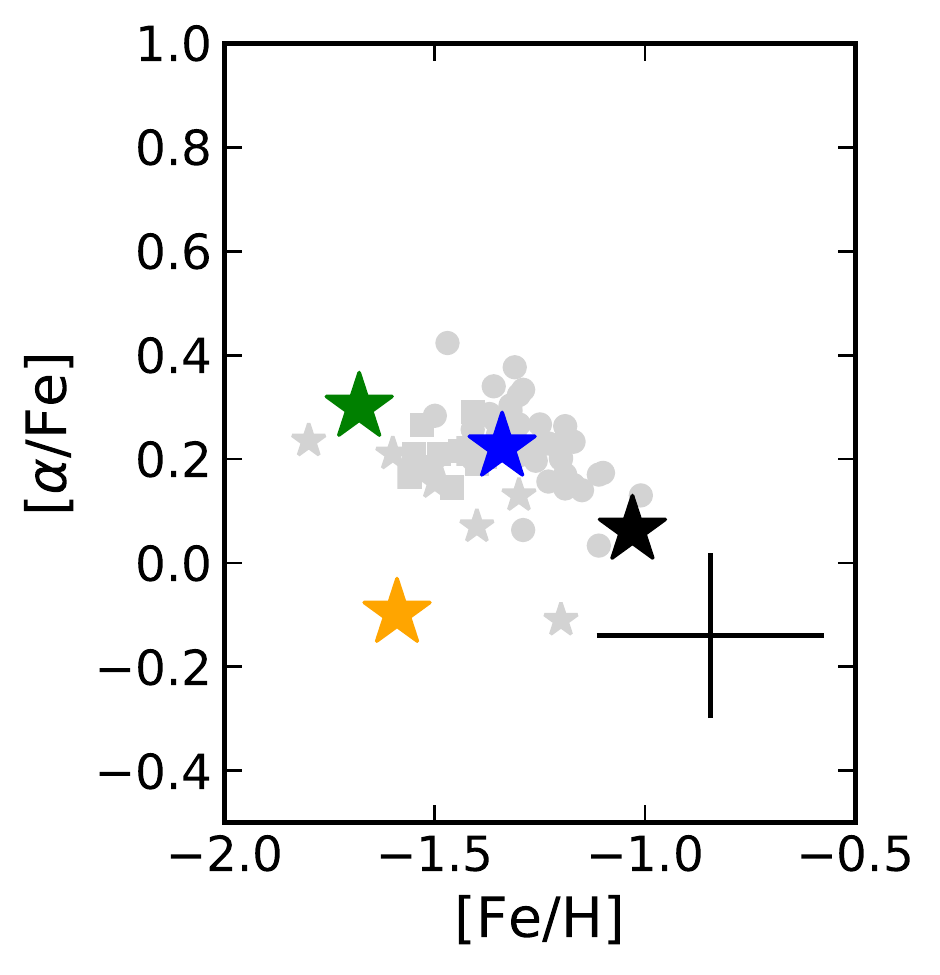}
\caption{Mean $\mathrm{[\alpha/Fe]}$ abundances as a function of metallicity for the four \gjoll stars and the stars in \glob. Literature abundances for stars in \glob are plotted in grey. Dots are abundances taken from \citet{meszaros2020}, squares from \citet{carretta2009b} and stars are from \citet{magurno2018}. Abundances from \citet{carretta2009b} are used for stars overlapping between the \citet{carretta2009b} and \citet{meszaros2020}. A representative error bar for the \gjoll abundances is shown in the lower right corner. \label{fig:alpha}}
\end{figure}

One of the most prominent features of stars born in GCs is the abundance correlations and anti-correlations found for specific element pairs. In Figure \ref{fig:abun} we plot the abundances of the four \gjoll stars along with abundances derived for stars in \glob from \citet{carretta2009b} and \citet{meszaros2020}. We inspect the specific abundance spaces of $\mathrm{[Al/Fe]}$ vs $\mathrm{[Mg/Fe]}$, $\mathrm{[Al/Fe]}$ vs $\mathrm{[Si/Fe]}$, $\mathrm{[Mg/Fe]}$ vs $\mathrm{[Si/Fe]}$,  and $\mathrm{[Al/Fe]}$ vs $\mathrm{[Na/Fe]}$, where correlations or anti-correlations are known to exist for GCs and especially for \glob. We also compare the mean $\alpha$ abundances of the \gjoll stars with the stars in \glob in Figure \ref{fig:alpha}, in this plot we have included data from the study of \citet{magurno2018} also. For the \gjoll stars, and for the stars from \citet{meszaros2020}, $\mathrm{[\alpha/Fe]}$ is calculated as $\mathrm{[\langle Mg,Si,Ca\rangle/Fe]}$. \citet{carretta2009b} did not derive Ca abundances for their stars, so here only Mg and Si are used for the mean $\alpha$ abundance, and for \citet{magurno2018} only Mg and Ca abundances are available. All abundances from the literature have been recalculated using the \cite{asplund2009} solar abundance scale. From these figures, it can be seen that the abundances derived for $Gaia$~31873905 and $Gaia$~29901421 (green and blue star, respectively) both display a good match to the abundances derived for the \glob stars. The abundances derived for $Gaia$~32541125 (black star), however, display a fairly poor match to the GC star abundances. The mismatch is driven primarily by this star's low upper limit on the Al abundance. Thus from a chemical point of view, $Gaia$~31873905 and $Gaia$~29901421 were likely stripped from \glob, while $Gaia$~32541125 is likely an unassociated field halo star. For $Gaia$~32591587, shown as a yellow star in Figures \ref{fig:abun} and \ref{fig:alpha}, we find a good match to \glob in the Al abundance, but the low $\alpha$-element abundances and high Na abundance found in this star is less of a perfect match to the \glob stars displayed in the figures. It should be noted that \citet{magurno2018} also finds $\mathrm{[Mg/Fe]} < 0$ and $\mathrm{[Ca/Fe]} < 0 $ for some of the \glob RRL stars they analyse (see Figure \ref{fig:alpha}), thus despite the offset in $\alpha$-element abundances from the sample of giant stars analysed by \citet{carretta2009b} and \citet{meszaros2020}, $Gaia$~32591587 is still likely to have been stripped from \glob.

\begin{figure}[hbt!]
\centering
\includegraphics[width=\linewidth]{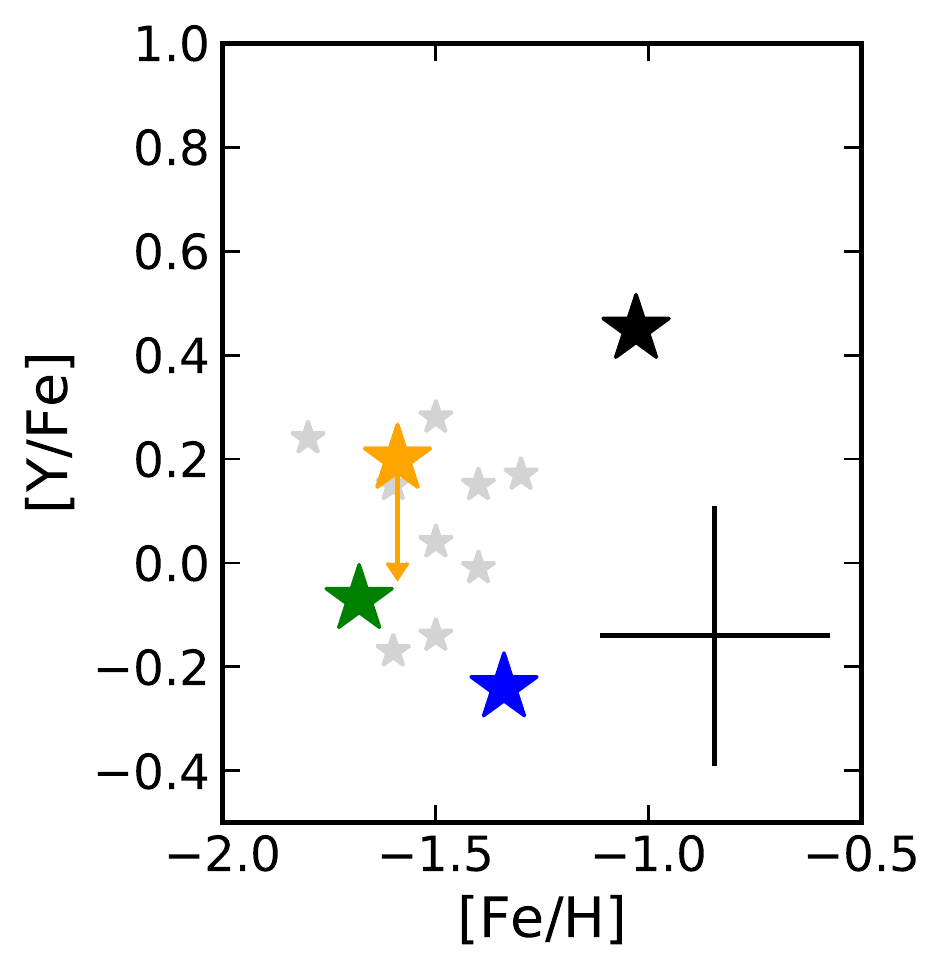}
\caption{Mean $\mathrm{[Y/Fe]}$ abundances as a function of metallicity for the four \gjoll stars and stars in \glob from \citet{magurno2018}. Symbols as in Figure \ref{fig:abun}.\label{fig:y}}
\end{figure}

A subset of GCs, including $\omega$ Cen that was tied to the Fimbulthul stream, show enhancement in elements produced in the slow neutron-capture process ($s$-process). This particular abundance signature revealed by high $\mathrm{[Y/Fe]}$ and $\mathrm{[Ba/Fe]}$ ratios played a central role in strengthening the case of a chemical match between $\omega$ Cen and the Fimbulthul stream. There are only a few reports of neutron-capture element abundances for stars in \glob. One of these is \citet{magurno2018}, who derived Y abundances for nine stars in \glob and found an average abundance of $\mathrm{[Y/Fe]}=0.08$ suggesting that \glob is not strongly enhanced in $s$-process elements. In Figure \ref{fig:y} we plot the $\mathrm{[Y/Fe]}$ for the \gjoll stars along with the data from \citet{magurno2018}.  Again we see that the abundances of $Gaia$~31873905 and $Gaia$~29901421 display the best match to the abundances of \glob.

\section{Summary} \label{sec:summary}
We present a chemo-dynamical analysis of stars in the northern stream \gjoll, which is likely the result of dynamical interaction between the GC \glob and the Milky Way \citep{riley2020}. The orbit of \glob in a Milky Way potential \citep{pricewhelan2017} overlaps with both endpoints of the stream. Using this orbit, we've identified 162 stars that are kinematic members of this system for spectroscopic follow-up, five of which were bright enough to obtain high-resolution spectroscopy for. 

Based on our measured radial velocities four of the five stars remain good member candidates, while $Gaia$~32589760 is not a member. We further present a detailed chemical analysis for the four stars with high resolution spectra $Gaia$~32541125, $Gaia$~31873905, $Gaia$~29901421, and $Gaia$~32591587, comparing the derived abundances to those of stars in \glob. The abundances of $Gaia$~31873905 and $Gaia$~29901421 match well with the previously determined abundance patterns in \glob, following the cluster's expected abundance correlations and anti-correlation, and showing low $\mathrm{[Y/Fe]}$ abundances. For $Gaia$~32591587, we find a partial match between the abundances of this star and the stars in \glob. Finally, a poor match is found for $Gaia$~32541125, which displays an abundance pattern more similar to a halo field star, notably with a low upper limit on the Al abundance and higher $\mathrm{[\alpha/Fe]}$ and neutron-capture element abundances. Based on this analysis, we conclude that $Gaia$~31873905, $Gaia$~29901421, and $Gaia$~32591587 were all stripped from \glob, demonstrating both a kinematic and chemical tie between \glob and the \gjoll stream. Our results highlight the power of and need for a combination of dynamical and detailed chemical analysis when linking streams to their parent object.

Spectroscopic followup of the fainter stars identified here as candidate members could further confirm this association. Additionally, we note that disrupting globular clusters typically form symmetric tidal tails. Searching along the projected future orbit of \glob, or further along the trailing orbit than the $l=180^\circ$ cutoff of the original \gjoll discovery, could also reveal more members and help further characterize this system.

\acknowledgements{We thank the referee for their timely report and insightful comments on the paper. AHR acknowledges support from a Texas A\&M University Merit Fellowship and an NSF Graduate Research Fellowship through Grant DGE-1746932. T.T.H., P.S.F., J.L.M., and L.E.S. acknowledge generous support from the George P. and Cynthia Woods Institute for Fundamental Physics and Astronomy at Texas A\&M University. This research made extensive use of the SIMBAD database operated at CDS, Straasburg, France \cite{wenger2000}, \href{https://arxiv.org/}{arXiv.org}, and NASA's Astrophysics Data System for bibliographic information.}

\facilities{Smith}
\software{Astropy \citep{Astropy:13, Astropy:18}, ATLAS9 \citep{castelli2003}, IRAF \citep{tody1986,tody1993}, gala \citep{gala, gala-zenodo}, Jupyter \citep{jupyter}, linemake (https://github.com/vmplacco/linemake), Matplotlib \citep{matplotlib}, MOOG \citep{sneden1973,sobeck2011}, NumPy \citep{numpy}, Pandas \citep{pandas}, SciPy \citep{scipy}, sfdmap (https://github.com/kbarbary/sfdmap)}

\end{document}

%% file: authors.tex
\author[0000-0001-6154-8983]{T.~T.~Hansen}
\altaffiliation{Mitchell Astronomy Fellow}
\affil{George P.~and Cynthia Woods Mitchell Institute for Fundamental Physics and Astronomy, Texas A\&M University, College Station, TX 77843, USA}
\affil{Department of Physics and Astronomy, Texas A\&M University, College Station, TX 77843, USA}

\author[0000-0001-5805-5766]{A.~H.~Riley}
\altaffiliation{Code for this work is available \href{https://github.com/ahriley/gjoll-ngc3201}{on Github}}.
\affil{George P.~and Cynthia Woods Mitchell Institute for Fundamental Physics and Astronomy, Texas A\&M University, College Station, TX 77843, USA}
\affil{Department of Physics and Astronomy, Texas A\&M University, College Station, TX 77843, USA}

\author{L.~E.~Strigari}
\affil{George P.~and Cynthia Woods Mitchell Institute for Fundamental Physics and Astronomy, Texas A\&M University, College Station, TX 77843, USA}
\affil{Department of Physics and Astronomy, Texas A\&M University, College Station, TX 77843, USA}

\author{J.~L.~Marshall}
\affil{George P.~and Cynthia Woods Mitchell Institute for Fundamental Physics and Astronomy, Texas A\&M University, College Station, TX 77843, USA}
\affil{Department of Physics and Astronomy, Texas A\&M University, College Station, TX 77843, USA}

\author{P.~S.~Ferguson}
\affil{George P.~and Cynthia Woods Mitchell Institute for Fundamental Physics and Astronomy, Texas A\&M University, College Station, TX 77843, USA}
\affil{Department of Physics and Astronomy, Texas A\&M University, College Station, TX 77843, USA}

\author{J.~Zepeda}
\affiliation{Department of Physics, University of Notre Dame, Notre Dame, IN 46556, USA}
\affiliation{Joint Institute for Nuclear Astrophysics -- Center for the Evolution of the Elements (JINA-CEE), USA}

\author{C.~Sneden}
\affiliation{Department of Astronomy and McDonald Observatory, The University of Texas, Austin, TX 78712, USA}

%% file: tables/candidates.tex
3258976074166599680 & $63.326$ & $1.827$ & 14.388 & 0.862 & $0.34 \pm 0.04$ & $18.04 \pm 0.06$ & $-19.95 \pm 0.04$ & $112.10 \pm 3.00$ &  \\
3259158764894232192 & $63.739$ & $2.568$ & 13.458 & 0.594 & $0.23 \pm 0.03$ & $19.43 \pm 0.06$ & $-20.15 \pm 0.03$ & $-150.60 \pm 3.50$ &  \\
3254112556278356608 & $65.582$ & $-0.864$ & 14.036 & 0.978 & $0.45 \pm 0.04$ & $15.64 \pm 0.06$ & $-16.47 \pm 0.03$ & $-79.50 \pm 0.90$ & N \\
3229373063616887936 & $69.793$ & $-1.536$ & 15.758 & 0.843 & $0.64 \pm 0.05$ & $19.45 \pm 0.09$ & $-19.57 \pm 0.06$ & $-33.17 \pm 0.65$* & N \\
3200439105894310272 & $72.085$ & $-5.176$ & 18.732 & 0.865 & $0.29 \pm 0.24$ & $20.91 \pm 0.41$ & $-21.38 \pm 0.34$ & $-15.18 \pm 4.38$* & Y \\
3188058536245928576 & $72.848$ & $-6.759$ & 16.407 & 0.650 & $0.31 \pm 0.06$ & $22.79 \pm 0.11$ & $-23.69 \pm 0.09$ & $0.81 \pm 3.18$* & N \\
3187390548572555136 & $74.627$ & $-6.423$ & 13.145 & 1.036 & $0.23 \pm 0.02$ & $22.12 \pm 0.04$ & $-23.09 \pm 0.03$ & $-10.30 \pm 0.60$ & Y \\
2990142148280216960 & $78.038$ & $-10.897$ & 13.100 & 0.528 & $0.23 \pm 0.02$ & $23.45 \pm 0.03$ & $-24.00 \pm 0.03$ & $7.20 \pm 0.60$ &  \\
2985543956292701312 & $82.104$ & $-13.340$ & 15.748 & 0.771 & $0.24 \pm 0.05$ & $23.96 \pm 0.08$ & $-24.21 \pm 0.08$ & $74.41 \pm 1.51$* & Y \\
\hline 
3278102525607689984 & $56.825$ & $7.715$ & 17.012 & 0.650 & $0.06 \pm 0.12$ & $15.86 \pm 0.26$ & $-15.34 \pm 0.19$ &  &  \\
3277467660721519616 & $57.617$ & $8.032$ & 18.703 & 0.839 & $-0.39 \pm 0.35$ & $16.00 \pm 0.68$ & $-15.85 \pm 0.54$ &  &  \\
3301522634377901952 & $57.851$ & $8.542$ & 17.992 & 0.698 & $0.14 \pm 0.24$ & $16.16 \pm 0.47$ & $-15.18 \pm 0.35$ &  &  \\
3301425396318555008 & $58.323$ & $7.947$ & 17.815 & 0.634 & $0.39 \pm 0.22$ & $16.72 \pm 0.41$ & $-17.77 \pm 0.28$ &  &  \\
3273949498390088448 & $58.377$ & $6.254$ & 16.420 & 0.646 & $0.23 \pm 0.11$ & $15.01 \pm 0.19$ & $-15.69 \pm 0.14$ &  &  \\
\vdots & \vdots & \vdots & \vdots & \vdots & \vdots & \vdots & \vdots & \vdots & \vdots \\

%% file: tables/RRL.tab
\begin{deluxetable}{rrrr}
\tablecaption{Detailed observing log for RRL stars.\label{tab:RRL}}
\tablewidth{0pt}
\tabletypesize{\scriptsize}
\tablehead{\colhead{HJD} & \colhead{$t_{exp}$} & \colhead{$V_{hel}$} & \colhead{Phase}\\
\colhead{} &\colhead{sec} & \colhead{Km~s$^{-1}$} &\colhead{}}
\startdata
\multicolumn{4}{c}{$Gaia$~29901421}\\
\hline
2458802.84375  &  2400  &   7.4$\pm 0.6$ & 0.937 \\
2458802.87305  &  2400  &   7.3$\pm 0.8$ & 0.988 \\
2458802.90234  &  2400  &   8.1$\pm 0.8$ & 0.039 \\
2458866.66113  &  1800  &  20.1$\pm 1.1$ & 0.142 \\
2458866.67969  &  962   &  23.2$\pm 1.6$ & 0.181 \\
\hline
\multicolumn{4}{c}{$Gaia$~32591587}\\
\hline
2458832.75000 &2400 &$-154.3\pm 1.2$ & 0.911 \\
2458888.57422 & 900 &$-144.3\pm 1.3$ & 0.170 \\
2458888.58594 & 900 &$-142.3\pm 1.2$ & 0.192 \\
2458888.59766 & 900 &$-138.8\pm 1.6$ & 0.213 \\
2458888.60938 & 900 &$-136.8\pm 1.2$ & 0.234 \\
2458888.62109 & 900 &$-135.0\pm 0.9$ & 0.255 \\
2458888.63281 & 900 &$-130.3\pm 1.6$ & 0.276 \\
2458888.64453 & 900 &$-131.1\pm 1.3$ & 0.297 \\
2458888.65625 & 900 &$-130.6\pm 2.0$ & 0.318 \\
\enddata
\end{deluxetable}

%% file: tables/Fe_ew_stub.tab
$Gaia$~32541125  &  \ion{Fe}{1} &4067.97 &  3.209 & $-$0.53 &  97.1 & 3.5&   6.44\\
$Gaia$~32541125  &  \ion{Fe}{1} &4150.24 &  3.428 & $-$1.19 &  61.5 & 3.9&   6.66\\
$Gaia$~32541125  &  \ion{Fe}{1} &4173.92 &  0.989 & $-$3.29 &  68.6 & 3.4&   6.62\\
$Gaia$~32541125  &  \ion{Fe}{1} &4174.91 &  0.914 & $-$2.94 &  83.7 & 3.3&   6.57\\
$Gaia$~32541125  &  \ion{Fe}{1} &4476.07 &  3.686 & $-$0.34 &  85.5 & 4.0&   6.64\\
$Gaia$~32541125  &  \ion{Fe}{1} &4595.35 &  3.299 & $-$1.73 &  44.1 & 4.7&   6.69\\
$Gaia$~32541125  &  \ion{Fe}{1} &4630.12 &  2.277 & $-$2.58 &  41.7 & 3.6&   6.33\\
\multicolumn{1}{c}{\vdots} & \multicolumn{1}{l}{\phantom{0}\vdots} & \multicolumn{1}{c}{\vdots} & \multicolumn{1}{c}{\vdots} & \multicolumn{1}{c}{\vdots} & \multicolumn{1}{c}{\vdots} & \multicolumn{1}{c}{\vdots} \\
$Gaia$~32541125  &  \ion{Fe}{2} &4489.18 &  2.828 & $-$2.96 &  51.3 & 3.8&   6.76\\
$Gaia$~32541125  &  \ion{Fe}{2} &4520.22 &  2.807 & $-$2.65 &  48.5 & 3.7&   6.34\\
$Gaia$~32541125  &  \ion{Fe}{2} &4522.63 &  2.840 & $-$2.29 &  60.7 & 3.9&   6.36\\
$Gaia$~32541125  &  \ion{Fe}{2} &4534.16 &  2.856 & $-$3.28 &  34.9 & 3.4&   6.61\\
$Gaia$~32541125  &  \ion{Fe}{2} &4541.52 &  2.856 & $-$2.98 &  44.8 & 4.0&   6.60\\

%% file: tables/lineabun_stub.tab
CH         &  4300.000 &\nodata&\nodata& synth &7.40   &  synth &6.84   &\nodata&\nodata&\nodata& \nodata\\
\ion{Na}{1}&  5682.633 & 2.101 & $-$0.70 &\nodata&\nodata&  27.90 &5.00   &\nodata&\nodata&\nodata& \nodata\\
\ion{Na}{1}&  5688.205 & 2.103 & $-$0.45 & 78.43 &5.52   &  53.96 &5.10   &\nodata&\nodata&\nodata& \nodata\\
\ion{Na}{1}&  5895.924 & 0.000 & $-$0.18 &\nodata&\nodata&\nodata &\nodata&\nodata&\nodata& 171.10& 5.27\\ 
\ion{Mg}{1}&  4167.271 & 4.343 & $-$1.00 & 83.20 &6.31   &\nodata &\nodata& 27.80 &5.84&\nodata&\nodata\\
\ion{Mg}{1}&  4571.096 & 0.000 & $-$5.69 & 71.82 &6.38   & 109.20 &6.63   &\nodata&\nodata&\nodata&\nodata\\
\ion{Mg}{1}&  4702.991 & 4.343 & $-$0.67 &128.80 &6.27   & 136.40 &6.46	& 94.41 &6.42	&  67.60& 6.01\\

%% file: tables/error.tab
CH              & 0.14 & 0.06 & 0.01 & 0.05 & 0.21   \\
\ion{Na}{1}     & 0.06 & 0.02 & 0.02 & 0.00 & 0.07   \\
\ion{Mg}{1}     & 0.11 & 0.06 & 0.08 & 0.01 & 0.15   \\
\ion{Al}{1}     & 0.15 & 0.05 & 0.00 & 0.04 & 0.21   \\
\ion{Si}{1}     & 0.07 & 0.05 & 0.02 & 0.03 & 0.12   \\
\ion{K}{1}      & 0.12 & 0.05 & 0.13 & 0.01 & 0.18   \\
\ion{Ca}{1}     & 0.11 & 0.06 & 0.12 & 0.00 & 0.16   \\
\ion{Sc}{1}     & 0.02 & 0.13 & 0.06 & 0.04 & 0.10   \\
\ion{Ti}{2}     & 0.03 & 0.11 & 0.15 & 0.02 & 0.07   \\
\ion{V}{2}      & 0.02 & 0.11 & 0.09 & 0.02 & 0.05   \\
\ion{Cr}{1}     & 0.15 & 0.04 & 0.12 & 0.01 & 0.21   \\
\ion{Mn}{1}     & 0.11 & 0.02 & 0.05 & 0.01 & 0.14   \\
\ion{Fe}{1}     & 0.12 & 0.04 & 0.12 & 0.01 & 0.18   \\
\ion{Fe}{2}     & 0.01 & 0.13 & 0.08 & 0.02 & 0.05   \\
\ion{Co}{1}     & 0.16 & 0.02 & 0.14 & 0.00 & 0.23   \\
\ion{Ni}{1}     & 0.10 & 0.01 & 0.08 & 0.00 & 0.14   \\
\ion{Cu}{1}     & 0.11 & 0.01 & 0.01 & 0.00 & 0.12   \\
\ion{Zn}{1}     & 0.03 & 0.08 & 0.10 & 0.01 & 0.06   \\
\ion{Sr}{2}     & 0.14 & 0.08 & 0.07 & 0.04 & 0.21   \\
\ion{Y}{2}      & 0.03 & 0.10 & 0.10 & 0.01 & 0.06   \\
\ion{Zr}{2}     & 0.09 & 0.14 & 0.04 & 0.02 & 0.19   \\
\ion{Ba}{2}     & 0.04 & 0.11 & 0.11 & 0.01 & 0.08   \\
\ion{La}{2}     & 0.13 & 0.11 & 0.02 & 0.02 & 0.21   \\
\ion{Ce}{2}     & 0.06 & 0.12 & 0.03 & 0.04 & 0.15   \\
\ion{Nd}{2}     & 0.08 & 0.11 & 0.00 & 0.00 & 0.17   \\
\ion{Eu}{2}$^*$ & 0.09 & 0.08 & 0.01 & 0.02 & 0.15   \\

%% file: tables/error_RRL.tab
\ion{Fe}{1}     & 0.11 & 0.02 & 0.05 & 0.01 & 0.14   \\
\ion{Na}{1}     & 0.08 & 0.02 & 0.01 & 0.00 & 0.10   \\
\ion{Mg}{1}     & 0.09 & 0.01 & 0.09 & 0.01 & 0.14   \\
\ion{Al}{1}     & 0.18 & 0.04 & 0.12 & 0.06 & 0.27   \\
\ion{Si}{1}     & 0.23 & 0.03 & 0.00 & 0.04 & 0.28   \\
\ion{K}{1}      & 0.11 & 0.02 & 0.02 & 0.00 & 0.13   \\
\ion{Ca}{1}     & 0.10 & 0.02 & 0.04 & 0.00 & 0.12   \\
\ion{Sc}{1}     & 0.03 & 0.08 & 0.00 & 0.02 & 0.11   \\
\ion{Ti}{2}     & 0.06 & 0.11 & 0.09 & 0.00 & 0.17   \\
\ion{V}{2}      & 0.06 & 0.08 & 0.04 & 0.02 & 0.14   \\
\ion{Cr}{1}     & 0.14 & 0.01 & 0.09 & 0.00 & 0.17   \\
\ion{Fe}{2}     & 0.02 & 0.10 & 0.08 & 0.00 & 0.12   \\
\ion{Ni}{1}     & 0.10 & 0.02 & 0.01 & 0.00 & 0.12   \\
\ion{Cu}{1}     & 0.12 & 0.04 & 0.00 & 0.01 & 0.16   \\
\ion{Zn}{1}     & 0.11 & 0.01 & 0.01 & 0.00 & 0.12   \\
\ion{Sr}{2}     & 0.12 & 0.13 & 0.20 & 0.01 & 0.28   \\
\ion{Y}{2}      & 0.01 & 0.11 & 0.04 & 0.01 & 0.11   \\
\ion{Zr}{2}     & 0.08 & 0.07 & 0.02 & 0.01 & 0.15   \\
\ion{Ba}{2}     & 0.11 & 0.06 & 0.04 & 0.04 & 0.19   \\

%% file: tables/abun1.tab
CH          &$+$6.84 & 1 & 0.20  &$-$1.59&0.29 &$-$0.25 &0.32 	&\nodata &  &\nodata&\nodata&\nodata&\nodata&\nodata\\
\ion{Na}{1} &$+$5.00 & 2 & 0.14  &$-$1.24&0.16 &$+$0.10 &0.27 	&$+$4.32 &  & 0.20  &$-$1.92& 0.22  &$-$0.24& 0.27  \\
\ion{Mg}{1} &$+$6.53 & 4 & 0.09  &$-$1.07&0.18 &$+$0.27 &0.23 	&$+$6.25 & 5& 0.25  &$-$1.36& 0.28  &$+$0.33& 0.31  \\
\ion{Al}{1} &$+$5.18 & 3 & 0.32  &$-$1.27&0.38 &$+$0.07 &0.40 	&$+$4.77 & 1& 0.20  &$-$1.68& 0.34  &$+$0.00& 0.29  \\
\ion{Si}{1} &$+$6.27 & 1 & 0.20  &$-$1.24&0.23 &$+$0.10 &0.31 	&$+$6.13 & 1& 0.20  &$-$1.38& 0.34  &$+$0.30& 0.30  \\
\ion{K}{1}  &$+$4.26 & 1 & 0.20  &$-$0.77&0.27 &$+$0.57 &0.28 	&$+$4.03 & 1& 0.20  &$-$1.00& 0.24  &$+$0.68& 0.26  \\
\ion{Ca}{1} &$+$5.30 &16 & 0.19  &$-$1.05&0.25 &$+$0.30 &0.28 	&$+$4.93 &13& 0.23  &$-$1.41& 0.26  &$+$0.27& 0.29  \\
\ion{Sc}{2} &$+$1.77 & 3 & 0.04  &$-$1.38&0.11 &$-$0.04 &0.26 	&$+$1.18 & 5& 0.15  &$-$1.97& 0.19  &$-$0.29& 0.24  \\
\ion{Ti}{2} &$+$3.79 &32 & 0.21  &$-$1.16&0.22 &$+$0.18 &0.29 	&$+$3.51 &19& 0.24  &$-$1.44& 0.29  &$+$0.24& 0.30  \\
\ion{V}{2}  &$+$2.76 & 1 & 0.20  &$-$1.17&0.21 &$+$0.17 &0.31 	&$+$2.59 & 1& 0.20  &$-$1.34& 0.24  &$+$0.34& 0.26  \\
\ion{Cr}{1} &$+$4.22 &18 & 0.15  &$-$1.42&0.27 &$-$0.08 &0.26 	&$+$3.96 & 4& 0.35  &$-$1.68& 0.39  &$+$0.00& 0.39  \\
\ion{Mn}{1} &$+$3.75 & 3 & 0.04  &$-$1.68&0.14 &$-$0.34 &0.21 	&\nodata &  &\nodata&\nodata&\nodata&\nodata&\nodata\\
\ion{Fe}{1} &$+$6.16 &102& 0.11  &$-$1.34&0.27 &\nodata &\nodata&$+$5.82 &47& 0.16  &$-$1.68& 0.22  &\nodata&\nodata\\
\ion{Fe}{2} &$+$6.16 &19 & 0.07  &$-$1.34&0.09 &\nodata &\nodata&$+$5.82 &18& 0.11  &$-$1.68& 0.16  &\nodata&\nodata\\
\ion{Co}{1} &$+$3.30 & 1 & 0.20  &$-$1.69&0.30 &$-$0.35 &0.29 	&\nodata &  &\nodata&\nodata&\nodata&\nodata&\nodata\\
\ion{Ni}{1} &$+$4.81 &13 & 0.13  &$-$1.41&0.19 &$-$0.07 &0.24 	&$+$4.46 & 1& 0.20  &$-$1.76& 0.23  &$-$0.08& 0.27  \\
\ion{Cu}{1} &$+$1.92 & 1 & 0.20  &$-$2.27&0.23 &$-$0.93 &0.29 	&$+$2.89 & 1& 0.20  &$-$1.30& 0.26  &$+$0.38& 0.27  \\
\ion{Zn}{1} &$+$3.39 & 2 & 0.20  &$-$1.17&0.21 &$+$0.17 &0.30 	&$+$3.40 & 1& 0.20  &$-$1.16& 0.23  &$+$0.52& 0.27  \\
\ion{Sr}{2} &$+$1.68 & 2 & 0.21  &$-$1.19&0.30 &$+$0.15 &0.31 	&$+$1.22 & 2& 0.20  &$-$1.66& 0.35  &$+$0.03& 0.30  \\
\ion{Y}{2}  &$+$0.63 & 4 & 0.11  &$-$1.58&0.13 &$-$0.24 &0.25 	&$+$0.46 & 1& 0.20  &$-$1.75& 0.23  &$-$0.07& 0.27  \\
\ion{Zr}{2} &$+$1.24 & 1 & 0.20  &$-$1.34&0.27 &$+$0.00 &0.34 	&$+$0.95 & 1& 0.20  &$-$1.63& 0.25  &$+$0.05& 0.27  \\
\ion{Ba}{2} &$+$0.79 & 2 & 0.20  &$-$1.39&0.21 &$-$0.05 &0.30 	&$+$0.13 & 2& 0.20  &$-$2.05& 0.27  &$-$0.37& 0.27  \\
\ion{La}{2} &$-$0.39 & 1 & 0.20  &$-$1.49&0.29 &$-$0.15 &0.33 	&\nodata &  &\nodata&\nodata&\nodata&\nodata&\nodata\\
\ion{Ce}{2} &$+$0.29 & 2 & 0.20  &$-$1.29&0.25 &$+$0.05 &0.33 	&\nodata &  &\nodata&\nodata&\nodata&\nodata&\nodata\\
\ion{Nd}{2} &$+$0.28 & 2 & 0.20  &$-$1.14&0.26 &$+$0.20 &0.35 	&\nodata &  &\nodata&\nodata&\nodata&\nodata&\nodata\\
\ion{Eu}{2} &\nodata &   &\nodata&\nodata&\nodata&\nodata&\nodata&\nodata &  &\nodata&\nodata&\nodata&\nodata&\nodata\\

%% file: tables/abun2.tab
CH          &\nodata   &  &\nodata&\nodata&\nodata&\nodata&\nodata  & $+$7.40 & 1& 0.20  &$-$1.03 &0.28   & $+$0.00 &0.30\\   
\ion{Na}{1} & $+$5.27  &1 & 0.20  &$-$0.97&0.22& $+$0.62&0.29       & $+$5.52 & 1& 0.20  &$-$0.72 &0.21   & $+$0.31 &0.29\\   
\ion{Mg}{1} & $+$5.87  &4 & 0.23  &$-$1.73&0.27& $-$0.14&0.32       & $+$6.50 & 5& 0.28  &$-$1.10 &0.32   & $-$0.07 &0.33\\   
\ion{Al}{1} & $+$4.86  &1 & 0.20  &$-$1.59&0.33& $+$0.00&0.32       &$<+$4.42 &  &\nodata&$<-$2.03&\nodata&$<-$1.00 &\nodata\\
\ion{Si}{1} & $+$5.92  &1 & 0.20  &$-$1.59&0.34& $+$0.00&0.33       &         & 3& 0.20  &$-$0.65 &0.23   & $+$0.38 &0.29\\   
\ion{K}{1}  & $+$2.76  &1 & 0.20  &$-$2.27&0.24& $-$0.68&0.29       & $+$4.70 & 2& 0.20  &$-$0.76 &0.27   & $+$0.70 &0.26\\   
\ion{Ca}{1} & $+$4.60  &9 & 0.31  &$-$1.74&0.33& $-$0.15&0.37       & $+$5.52 &15& 0.25  &$-$0.82 &0.30   & $+$0.21 &0.30\\   
\ion{Sc}{2} & $+$1.62  &4 & 0.08  &$-$1.53&0.14& $+$0.06&0.24       & $+$2.36 & 3& 0.17  &$-$0.79 &0.19   & $+$0.24 &0.29\\   
\ion{Ti}{2} & $+$3.45  &20& 0.27  &$-$1.50&0.32& $+$0.09&0.34       & $+$4.01 &23& 0.23  &$-$0.94 &0.24   & $+$0.09 &0.29\\   
\ion{V}{2}  & \nodata  &  &\nodata&\nodata&\nodata&\nodata&\nodata  & \nodata &  &\nodata&\nodata &\nodata& \nodata &\nodata\\
\ion{Cr}{1} & $+$3.99  &4 & 0.26  &$-$1.68&0.31& $-$0.09&0.34       & $+$4.30 & 8& 0.18  &$-$1.34 &0.27   & $-$0.31 &0.25\\   
\ion{Mn}{1} &$<+$3.84  &  &\nodata&$<-$1.59&\nodata&$<+$0.00&\nodata& $+$3.94 & 3& 0.38  &$-$1.49 &0.40   & $-$0.46 &0.42\\   
\ion{Fe}{1} &$+$5.91   &27&0.21   &$-$1.59 &0.25&\nodata&\nodata    & $+$6.47 &77& 0.17  &$-$1.03 &0.25   & \nodata &\nodata\\
\ion{Fe}{2} &$+$5.92   &14&0.18   &$-$1.58 &0.21&\nodata&\nodata    & $+$6.46 &15& 0.17  &$-$1.03 &0.28   & \nodata &\nodata\\
\ion{Co}{1} &$<+$3.40  &  &\nodata&$<-$1.59&\nodata&$<+$0.00&\nodata& $+$3.80 & 1& 0.20  &$-$1.19 &0.30   & $-$0.16 &0.27\\   
\ion{Ni}{1} & $+$4.45  &7 & 0.23  &$-$1.77&0.26& $-$0.18&0.31       & $+$5.19 & 8& 0.20  &$-$1.03 &0.24   & $+$0.00 &0.27\\   
\ion{Cu}{1} & $+$3.10  &  &\nodata&$<-$1.09&\nodata&$<+$0.50&\nodata& $+$2.94 & 1& 0.20  &$-$1.25 &0.23   & $-$0.22 &0.27\\   
\ion{Zn}{1} & $+$3.47  &  &\nodata&$<-$1.09&\nodata&$<+$0.50&\nodata& $+$4.09 & 1& 0.20  &$-$0.47 &0.21   & $+$0.56 &0.28\\   
\ion{Sr}{2} & $+$0.89  &2 & 0.30  &$-$1.98&0.26& $-$0.39&0.31       &$<+$1.84 &  &\nodata&$<-$1.03&\nodata&$<+$0.00 &\nodata\\
\ion{Y}{2}  & $+$0.82  &  &\nodata&$<-$1.39&\nodata&$<+$0.20&\nodata& $+$1.63 & 2& 0.20  &$-$0.58 &0.21   & $+$0.45 &0.28\\   
\ion{Zr}{2} & \nodata  &  &\nodata&\nodata&\nodata&\nodata&\nodata  & $+$2.05 & 1& 0.20  &$-$0.53 &0.27   & $+$0.50 &0.32\\   
\ion{Ba}{2} & $+$0.35  &2 & 0.20  &$-$1.83&0.27& $-$0.24&0.30       & $+$1.79 & 2& 0.20  &$-$0.40 &0.21   & $+$0.64 &0.28\\   
\ion{La}{2} & \nodata  &  &\nodata&\nodata&\nodata& \nodata&\nodata &\nodata  &  &\nodata&\nodata &\nodata&\nodata  &\nodata\\
\ion{Ce}{2} & \nodata  &  &\nodata&\nodata&\nodata& \nodata&\nodata & $+$1.30 & 2& 0.21  &$-$0.28 &0.29   & $+$0.75 &0.32\\   
\ion{Nd}{2} & \nodata  &  &\nodata&\nodata&\nodata& \nodata&\nodata & $+$1.19 & 1& 0.20  &$-$0.23 &0.26   & $+$0.80 &0.33\\   
\ion{Eu}{2} &$<-$0.57  &  &\nodata&$<-$1.09&\nodata&$<+$0.50&\nodata& $-$0.01 & 1& 0.20  &$-$0.53 &0.25   & $+$0.50 &0.31\\